\documentclass[transmag]{IEEEtran}

\usepackage{latexsym}
\usepackage{graphicx}
\usepackage{amsfonts,amssymb,amsmath}
\usepackage[nolist,nohyperlinks]{acronym}
\usepackage[numbers,sort&compress]{natbib}
\usepackage{multirow}
\usepackage{color}
\usepackage[dvipsnames]{xcolor}
\usepackage{mathtools}
\usepackage{bm}
\usepackage{algorithm2e}
\usepackage{mathabx}
\DeclarePairedDelimiter\ceil{\lceil}{\rceil}

\def\BibTeX{{\rm B\kern-.05em{\sc i\kern-.025em b}\kern-.08em T\kern-.1667em\lower.7ex\hbox{E}\kern-.125emX}}

\begin{acronym}
\acro{MIMO}{multiple-input multiple-output}
\acro{LIS}{large intelligent surface}
\acro{RIS}{reconfigurable intelligent surface}
\acro{DL}{Deep Learning}
\acro{CSI}{channel state information}
\acro{5G}{5th generation of wireless networks}
\acro{6G}{6th generation of wireless networks}
\acro{LoS}{line-of-sight}
\acro{LRD}{local reachability density}
\acro{FC}{fully connected layer}
\acro{PDF}{probability density function}
\acro{CDF}{cumulative distribution function}
\acro{SVM}{support vector machine}
\acro{SNR}{signal-to-noise ratio}
\acro{ML}{machine learning}
\acro{LOF}{local outlier factor}
\acro{DAE}{denoising autoencoder network}
\acrodefplural{DAE}[DAEs]{denoising autoencoder networks}
\acro{GLRT}{generalized likelihood ratio test}
\acro{MSE}{mean squared error}
\acro{HMIMO}{holographic MIMO surfaces}
\acro{RSS}{Received Signal Strength}
\acro{RTI}{radio tomographic images}
\end{acronym}

\ifCLASSOPTIONcompsoc
    \usepackage[caption=false, font=normalsize, labelfont=sf, textfont=sf]{subfig}
\else
\usepackage[caption=false, font=footnotesize]{subfig}
\fi

\makeatletter
\newcases{mycases}{\;}{%
   $\m@th\displaystyle{##}$}{$\m@th\displaystyle{##}$\hfil}{.}{.}
\makeatother

\SetKwInput{kwTrain}{Training phase}
\SetKwInput{kwEval}{Evaluation phase}

\begin{document}
%
\title{{Assessing Wireless Sensing Potential with  Large Intelligent Surfaces}}

\author{Cristian J. Vaca-Rubio, Pablo Ramirez-Espinosa, Kimmo Kansanen,\\ Zheng-Hua Tan, Elisabeth de Carvalho and Petar Popovski
\thanks{This work has been submitted to IEEE for possible publication. Copyright
may be transferred without notice, after which this version may no longer be
accessible. A preliminary version of this work has been accepted to 15-th EAI International Conference on Cognitive Radio Oriented Wireless Networks (CROWNCOM 2020) \cite{vaca2020primer}.}
\thanks{This project has received funding from the European Union’s Horizon 2020 research and innovation programme under the Marie Sklodowska-Curie grant agreement No 813999.}
\thanks{C. J. Vaca-Rubio, P. Ramirez-Espinosa, Z. H. Tan, E. de Carvalho and P. Popovski are with Department of Electronic Systems, Aalborg University, Denmark (email: \{cjvr, pres, zt, edc, petarp\}@es.aau.dk).}
\thanks{K. Kansanen is with Norwegian University of Science and Technology, Trondheim, Norway (email: kimmo.kansanen@ntnu.no). At the time of initiation of this work he was with Aalborg University.}}


\IEEEtitleabstractindextext{\begin{abstract}
Sensing capability 
is one of the most highlighted new feature
of  future 6G wireless networks. This paper addresses the sensing potential of Large Intelligent Surfaces (LIS) in an exemplary Industry 4.0 scenario. Besides the attention received by LIS in terms of communication aspects, it can offer a high-resolution rendering of the propagation environment. This is because, in an indoor setting, it can be placed in proximity to the sensed phenomena, while the high resolution is offered by densely spaced tiny antennas deployed over a large area. By treating an LIS as a radio image of the environment relying on the received signal power, we develop techniques to sense the environment, 
by leveraging the tools of image processing and machine learning. Once a radio image is obtained, a Denoising Autoencoder (DAE) network can be used for constructing a super-resolution image leading to sensing advantages not available in traditional sensing systems. Also, we derive a statistical test based on the Generalized Likelihood Ratio (GLRT) as a benchmark for the machine learning solution. We test these methods for a scenario where we need to detect whether an industrial robot deviates from a predefined route. The results show that the LIS-based sensing offers high precision and has a high application potential in indoor industrial environments. 
\end{abstract}

}

\maketitle
%
%
%
\section{Introduction}
\label{Introduction}

\IEEEPARstart{M}{assive} \ac{MIMO} is {one of the essential technologies} in the \ac{5G}
\cite{andrews2014will}.
Compared with traditional multiuser \ac{MIMO} systems, the base station of a massive MIMO system is equipped with a large number of antennas, aiming to further increase spectral efficiency \cite{larsson2014massive}. Looking towards the \ac{6G}, there are some significant breakthroughs on the design of reprogramable metamaterials \cite{shlezinger2021dynamic}, giving raise to new concepts such as \textit{\ac{HMIMO}} \cite{huang2020holographic}, \textit{\ac{LIS}} \cite{Hu2018} and \textit{\ac{RIS}} \cite{basar2019wireless}. While \ac{HMIMO} and \ac{LIS} originally refer to the use of continuous radiating surfaces where the received electromagnetic field is recorded and ultimately reconstructed \cite{Bjornson2019_Next}, in practice 
 an \ac{LIS} is envisioned and regarded as a collection of closely spaced tiny antenna elements. On the other hand, \ac{RIS} are composed by small passive reflectors embedded in a surface, allowing to arbitrarily modify the phase of the impinging electromagnetic waves and thus enabling a smart control of the propagation environment \cite{di2020smart,alexandropoulos2020reconfigurable,di2019smart}.
 
 The performance analysis of \ac{LIS} and \ac{RIS} assisted systems has attracted considerable attention in the recent years, and many works have studied the applicability of these technologies. For instance, the use of \ac{RIS} to control the signals propagation has been analyzed in the context of communications through the so-called passive beamforming \cite{wu2019intelligent, Huang2019, badiu2019communication, basar2019wireless}, in location and positioning systems \cite{wymeersch2019radio, Alghamdi2020, he2020large}, and in physical layer security \cite{Shen2019, Sanchez2020}. {The combination of \ac{DL} and \ac{RIS} elements efficient reconfiguration has also been studied in \cite{huang2019indoor}}. In turn, \ac{LIS} are considered as a natural extension of massive \ac{MIMO}, and their potential for data transmission \cite{Dardari2019, Hu2018} and positioning \cite{Hu2018Positioning, hu2017cramer} has been also addressed.
However, the potential of \ac{LIS} could go beyond communications applications. Indeed, such large surfaces contain many antennas that can be used as sensors of the environment based on the \ac{CSI}.

Sensing strategies based on electromagnetic signals have been thoroughly addressed in the literature in different ways, and applied to a wide range of applications. For instance, in \cite{wang2016rt}, a real-time fall detection system is proposed through the analysis of the communication signals produced by active users, whilst the authors in \cite{pu2013whole} use Doppler shifts for gesture recognition. Radar-like sensing solutions are also available for user tracking \cite{zhao2013radio} and real-time breath monitoring \cite{adib2014real}, as well as sensing methods based on \ac{RTI} \cite{zhao2018through, wilson2010radio}. Interestingly, whilst some of these techniques resort solely on the amplitude (equivalently, power) of the receive signals \cite{zhao2013radio, wilson2010radio}, in the cases where sensing small scale variations is needed, the full \ac{CSI} (i.e., amplitude and phase of the impinging signals) is required \cite{zhao2018through, adib2014real}.Moreover, in terms of power-based radio maps generation, some indoor positioning strategies leverage the use of \ac{ML} solutions \cite{bozkurt2015comparative} based on the \ac{RSS} of several beforehand known anchors for localization purposes.

On a related note, \ac{ML} based approaches are gaining popularity in the context of massive \ac{MIMO}, mainly due to the inherent complexity of this type of systems and their sensitivity to hardware impairments and channel estimation errors. Hence, \ac{DL} techniques arise as a promising solution to deal with massive \ac{MIMO}, and several works have shown the advantages of \ac{ML} solutions in channel estimation and precoding \cite{joung2016machine, Demir2020, Ma2020, Huang2018, Huang2019, Wen2018, Luo2020}. Due to the even larger dimensions of the system in extra-large arrays, \ac{DL} may play a key role in exploiting complex patterns of information dependency between the transmitted signals. Also, the popularization of \ac{LIS} as a natural next step from massive \ac{MIMO} gives rise to larger arrays and more degrees of freedom, providing huge amount of data which can feed \ac{ML} algorithms. 

Despite all the available works dealing with beyond massive \ac{MIMO} and sensing, both topics have  been addressed {rather} separately. {This has motivated the present work, where the objective is to assess the potential of  the combined use of \ac{DL} algorithms and  large surfaces for the purpose of sensing the propagation environment.}
To that end, the received signal along with the \ac{LIS} is treated as a radio image of the propagation environment, giving raise to the use of image processing techniques to improve the performance of sensing systems beyond purely radio-based approaches. Also, we analyze the pros and cons of this image sensing proposal, comparing it to alternative solutions based on classical post-processing of the received radio signal. Specifically, the contributions of this work are summarized as follows:
\begin{itemize}
    \item We propose an image-based sensing technique based on the received signal power at each antenna element of an \ac{LIS}. These power samples are processed to generate a high resolution image of the propagation environment that can be used to feed \ac{ML} algorithms to sense large-scale events. The usage of received signal power would lead to simple deployments, since there is no need of coherent receivers. 
    \item A \ac{ML} algorithm, based on transfer learning and \ac{LOF}, is defined to process the radio images generated by the \ac{LIS} in order to detect anomalies over a predefined robot route.
    \item We show the advantage of representing the radio propagation environment as an image, allowing us to use a \ac{DAE} for augmenting image resolution and significantly increasing the performance of the system.
    \item We derive a statistical test, based on the classical \ac{GLRT}, to carry out the same sensing task, and perform a comparison with the \ac{ML} solution in terms of generality, performance and further potential applications.
\end{itemize}

 To evaluate the capabilities for sensing of \ac{LIS}, we consider a simple problem of route anomaly detection in an indoor industrial scenario. Hence, we analyze the feasibility of this proposal to determine whether a robot has deviated from its predefined route, and compared it with the here derived statistical solution. 

The reminder of the paper is organized as follows. Section \ref{sec:HolographicSensing} introduces the concept of sensing based on radio images. Then, Section \ref{sec:ProblemFormulation} presents the problem of robot deviation detection in an industrial setup. The classical solution based on hypothesis testing is derived and characterized in Section \ref{sec:GLRT} and the proposed \ac{ML} algorithm is detailed in Section {\ref{sec:MLalgorithm}. With the \ac{ML} solution presented, the model validation is carried out in Section \ref{sec:Validation}, whilst simulated results are discussed in Section \ref{sec:Simulations}. Finally, some conclusions are drawn in Section \ref{sec:Conclusions}.
\begin{figure}[t]
\centering
\subfloat[LoS, noiseless. \label{fig:1a}]{\includegraphics[width=0.43\textwidth]{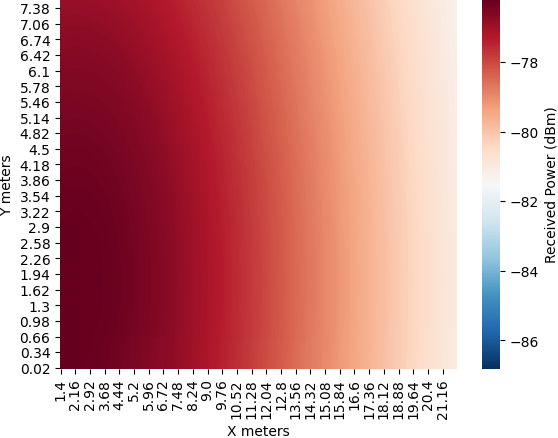}} \hfill
\subfloat[Real scenario, noiseless. \label{fig:1b}]{\includegraphics[width=0.43\textwidth]{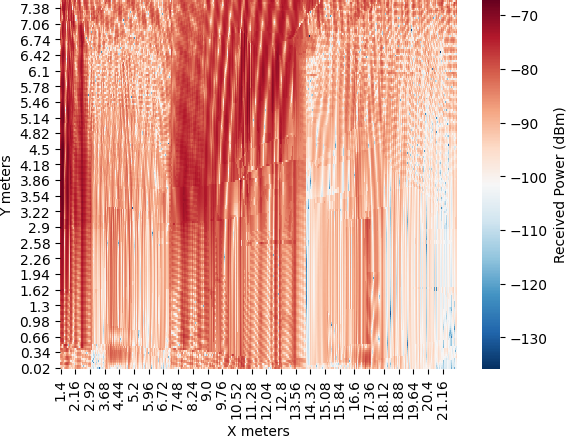}} \hfill

\caption{Radio images for \ac{LoS} and Industry scenarios.}
\label{fig:holImages}
\end{figure}

\section{Radio image-based LIS sensing}
\label{sec:HolographicSensing}


In a wireless context, a \ac{LIS} could be described as a structure which uses electromagnetic signals impinging in a determined scatterer in order to obtain a profile of the environment. That is, we can use the resulting signal of the superposition of all the involved paths that imping into every of the antenna elements conforming the surface. Then, the power of the resulting
superimposed signal is used to obtain a high resolution image of the propagation environment. Note the LIS elements are using the CSI information as envelope detectors, as no phase estimation is needed but the recieved signal power. Using this approach, the complexity of the multipath propagation is reduced to using information represented as an image. This provides a twofold benefit: \emph{i)} the massive number of elements that compose the \ac{LIS} leads to an accurate environment sensing (i.e. high resolution image), and \emph{ii)} it allows the use of computer vision algorithms and \ac{ML} techniques to deal with the resulting images. 

As an illustrative example, an LIS is deployed in a wall along a $22 \times 8$ m physical aperture, containing antenna elements separated $\frac{\lambda}{2}$ cm while an arbitrary robot is transmitting a sensing signal. Fig. \ref{fig:holImages} shows the LIS radio images obtained from different propagation environments under this setup. Specifically, Fig. \ref{fig:1a} corresponds to an \ac{LoS} propagation (no scatterers), whilst Fig. \ref{fig:1b} is obtained from an industrial scenario with a rich scattering.  Note that, in the case in which different scatterers are placed, their position and shapes are captured by the \ac{LIS} and represented in the image. {Beyond that, LIS-based imaging does not need of a previous calibration period as well as no scatterers need to be modelled to be captured in the radio image, contrary to other wireless image reconstruction techniques that rely on the received signal power, such as \ac{RTI} \cite{zhao2018through, wilson2010radio}.} To the best of the authors' knowledge, this is the first time that \ac{LIS} image based wireless sensing is proposed in the literature.}

\section{{System Model and Problem Formulation}}
\label{sec:ProblemFormulation}

We stick to a simple baseline problem in order to analyze, for first time in the literature, the sensing potential of \ac{LIS}. To that end, let consider an industrial scenario where a robot is supposed to follow a fixed route. Assume that, due to arbitrary reasons, it might temporarily deviate from the predefined route and follow an alternative (undesired) trajectory. The goal is to be able to detect whether the robot is following the correct route or not, based on the sensing signal transmitted by the target device. 

In order to perform the detection, we assume that a \ac{LIS}, {consisting of $M$ antenna elements,
is placed along a wall.} The sensing problem reduces to determine, from the received signal at each of the {$M$} \ac{LIS} elements, whether the transmission has been made from a point at the desired route or from anomalous ones. Formally, if we define the correct trajectory as the set of points in space $\mathbf{P}_c\in\mathbb{R}^{N_p\times 3} = (\mathbf{p}_1, \dots, \mathbf{p}_{N_p})$, and the received complex signal from an arbitrary point, $\mathbf{p}_k\in\mathbb{R}^{1\times 3}$, as $\mathbf{y}_k\in\mathbf{C}^{M\times 1}$, then the problem reduces to estimating whether $\mathbf{p}_k \in \mathbf{P}_c$ based on $\mathbf{y}_k$. Note that this formulation can be generalized to any anomalous detection based on radio-waves in an arbitrary scenario. 

The complex baseband signal received at the \ac{LIS} from point $\mathbf{p}$ (the subindex is dropped for the sake of clarity in the notation) is given by  
\begin{equation}
    \label{eq:RecSignal}
	\mathbf{y} = \mathbf{h} x + \mathbf{n},
\end{equation}
with $x$ the transmitted (sensing) symbol, $\mathbf{h}\in\mathbb{C}^{M\times 1}$ the channel vector from point $\mathbf{p}$ to each antenna-element, and $\mathbf{n}\sim\mathcal{CN}_{M}(\mathbf{0},\sigma^2\mathbf{I}_M)$ the noise vector. Moreover, we consider a static scenario where the the channel $\mathbf{h}$ only depends on the user position, neglecting the impact of time variations. 

In order to reduce deployment costs, and because we are interested in sensing large scale variations, we consider the received signal amplitude (equivalently, power). This assumption may lead to cheaper and simpler system implementations, avoiding the necessity of performing coherent detection.

\section{Statistical Approach: Likelihood Ratio Test}
\label{sec:GLRT}

\subsection{Decision rule}

Let us consider that the system is able to obtain several samples from each point $\mathbf{p}_k$ belonging to the correct route $\mathbf{P}_c$ during a training phase. Then, once the system is trained, the problem can be tackled from a statistical viewpoint by performing a generalized hypothesis test, as shown throughout this section.

{To start with, }let us assume that the value of $\sigma^2$ in \eqref{eq:RecSignal} is perfectly known and, without loss of generality, that $x = 1$. Since we are considering only received powers, the signal at the output of the $i$-th antenna detector is given by 
\begin{equation}
    \label{eq:signal}
    w_{i} = \|y_{i}\|^2 = \|h_{i} + n_{i}\|^2,
\end{equation}
where $y_{i}$, $h_{i}$ and $n_{i}$ for $i=1,\dots,M$ are the elements of $\mathbf{y}$, $\mathbf{h}$ and $\mathbf{n}$, respectively. When conditioned on $h_{i}$, $w_{i}$ follows a Rician distribution (in power) \cite{Durgin2002}, and due to the statistical independence of the noise samples, the joint conditioned \ac{PDF} of the vector  $\mathbf{w} = (w_1, \dots, w_M)^T$ is given by the product of the individual \acp{PDF}.

Consider also that, during the training phase, $N_t$ samples of $\mathbf{w}_0$, namely $\mathbf{w}_{0,j}$ for $j = 1,\dots,N_t$, are obtained from a correct (trained) point $\mathbf{p}_0$. The samples $\mathbf{w}_{0,j}$ are then jointly Rician distributed with vector parameter\footnote{Note that, due to the circular symmetry of the noise, the distribution of $\mathbf{w}$ does not depend on the complex channel $h_i$ but on its squared modulus $g_i = \|h_i\|^2$.} $\mathbf{g}_0 = (\|h_{0,1}\|^2, \dots, \|h_{0,M}\|^2)^T$. Then, from $\mathbf{w}_{0,j}$, the system obtains an estimation $\widehat{\mathbf{g}}_0 = (\widehat{g}_{0,1}, \dots, \widehat{g}_{0,M})^T$ of $\mathbf{g}_0$. 

Once trained (evaluation phase), the \ac{LIS} receives another set of samples $\mathbf{w}_{k}$ for $k = 1,\dots,N_v$ from an arbitrary point $\mathbf{p}$. Therefore, the objective is, based on $\mathbf{w}_{0,j}$ and $\mathbf{w}_{k}$, to determine whether $\mathbf{p} = \mathbf{p}_0$ or not. To that end, we formulate a binary hypothesis test as
\begin{equation}
\begin{mycases}
    H_0\; : & \;\widehat{\mathbf{g}} = \widehat{\mathbf{g}}_0 \\
     H_1\; : &  \;\widehat{\mathbf{g}} \neq \widehat{\mathbf{g}}_0 \\
     \end{mycases},
\end{equation}
where $\widehat{\mathbf{g}} = (\widehat{g}_i,\dots, \widehat{g}_M)^T$ is the channel vector estimated from $\mathbf{w}_k$. The test is hence formulated based on the \ac{GLRT}, but replacing the knowledge of the null hypothesis by its estimated counterpart, i.e.,
%
\begin{align}
    \textrm{log}\Lambda =& \sum_{k=1}^{N_v}\sum_{i=1}^M \textrm{log}I_0\left(\frac{2\widehat{g}_{0,i}\sqrt{w_{i,k}}}{\sigma^2}\right)+\sum_{k=1}^{N_v}\sum_{i=1}^M \frac{\widehat{g}_{i}-\widehat{g}_{0,i}}{\sigma^2} \notag \\
    &- \sum_{k=1}^{N_v}\sum_{i=1}^M\textrm{log}I_0\left(\frac{2\widehat{g}_{i}\sqrt{w_{i,k}}}{\sigma^2}\right) \underset{H_1}{\overset{H_0}{\gtrless}} \eta, \label{eq:GLRTlog}
\end{align}
where $w_{i,k}$ denote the $i$-th entry of $\mathbf{w}_k$. Replacing the true value of $\mathbf{g}_0$ by its estimation introduces a non-negligible error in the test that has to be considered in the threshold design, as we will see in the following subsections. 

\subsection{Estimator for $\mathbf{g}$}

In conventional likelihood ratio tests, the estimation of the involved parameters is carried out through maximum likelihood. However, since in our problem the distribution of the received power signal $\mathbf{w}_k$ $\forall$ $k$ is a multivariate Rician, the maximum likelihood estimation implies solving a system of $M$ non-linear equations \cite{Sijbers2001}. This may lead to a considerable computational effort taking into account the large number of antennas ($M$) that characterizes the \ac{LIS}. To circumvent this issue, we proposed a suboptimal --- albeit unbiased --- estimator based on moments. 

Since $\mathbb{E}[\mathbf{n}\mathbf{n}^H] = \mathbf{I}_M$, the estimation of the channel at each antenna element can be solved separately. Then, we can estimate $g_i$ in both the training and evaluation phases as
\begin{align}
    \widehat{g}_{0,i} &= \frac{1}{N_t}\sum_{j=1}^{N_t} w_{0,i,j} - \sigma^2,  &
    \widehat{g}_{i} &= \frac{1}{N_v}\sum_{k=1}^{N_v} w_{i,k} - \sigma^2, \label{eq:gestim}
\end{align}
where $w_{0,i,j}$ are the elements of $\mathbf{w}_{0,j}$. It is easily proved that the estimators in \eqref{eq:gestim} are unbiased with normally distributed error for relatively large number of samples. 
%

\subsection{Threshold design}

Although the asymptotic properties of $\textrm{log}\Lambda$ have been well studied in the literature (see, e.g., \cite{Kay1998}), these general results are not valid in our case because \textit{i)} we are replacing the true value of $\mathbf{g}_0$ by its estimation, and \textit{ii)} we are using moment-based estimators instead of the optimal one. A more general result, which is the starting point of our derivation, is that the the limiting distribution of $-2\textrm{log}\Lambda$, under the null hypothesis, is given by \cite[eq. (4.3)]{Scott2007}
\begin{equation}
    \label{eq:LimitingLambda}
    -2\textrm{log}\left.\Lambda\right|_{H_0} \overset{p}{\rightarrow} (\widehat{\mathbf{g}}_0-\widehat{\mathbf{g}})^T N_v\mathbf{J}(\widehat{\mathbf{g}}_0-\widehat{\mathbf{g}}),
\end{equation}
where we have replaced $\mathbf{g}_0$ by $\widehat{\mathbf{g}}_0$. In \eqref{eq:LimitingLambda}, $\overset{p}{\rightarrow}$ stands for convergence in probability and $\mathbf{J}\in \mathbb{R}^{M\times M}$ is the Fisher information matrix of $\mathbf{w}_k$ with respect to $\mathbf{g}_0$ \cite{Idier2014}. In our case, $\mathbf{J}$ is a diagonal matrix whose $i$-th element is given by 

\begin{equation}
    \label{eq:Ji}
    J_i(g_i) = \frac{e^{-g_i/\sigma^2}}{\sigma^6g_i}\int_0^\infty w_ie^{-w_i/\sigma^2}\frac{I_1^2\left(\frac{2}{\sigma^2}g_i\sqrt{w_i}\right)}{I_0\left(\frac{2}{\sigma^2}g_i\sqrt{w_i}\right)}dw_i - \frac{1}{\sigma^4}.
\end{equation}

Eq. \eqref{eq:LimitingLambda} can be rewritten as
\begin{equation}
    -2\textrm{log}\left.\Lambda\right|_{H_0} \overset{p}{\rightarrow} (\bm{\epsilon}_0 - \bm{\epsilon})^T N_v\mathbf{J}(\bm{\epsilon}_0 - \bm{\epsilon}),
\end{equation}
where $\bm{\epsilon}_0 = (\epsilon_{0,1},\dots,\epsilon_{0,M})^T$ and $\bm{\epsilon} = (\epsilon_{1},\dots,\epsilon_{M})^T$ are the error vectors of estimators in \eqref{eq:gestim}. Note that both error vectors are Gaussian distributed, but they vary at very different time scales. The true channel $\mathbf{g}_0$ is estimated during the training phase, and thus the error $\bm{\epsilon}_0$, albeit random, remains constant during the whole evaluation phase until the system is retrained. In turn, each time the system evaluates a point, $\bm{\epsilon}$ takes a different (random) realization. With that in mind, we propose choosing $\eta$ based on a worst case design, i.e., we consider an estimation error $\bm{\epsilon}_0$ that overestimates the true error at $1-\alpha_0$ percent of the time. That is, 
\begin{equation}
    \epsilon_{0,i}' = F_{\epsilon_{0,i}}^{-1}(1-\alpha_0/2), \label{eq:Epsilon0}
\end{equation}
where $F_{\epsilon_{0,i}}$ stands for the \ac{CDF} of a Gaussian variable with zero mean and variance $\frac{\sigma^2}{N_t}(\sigma^2+2\widehat{g}_{0,i})$. Note that we have replaced the true channel value by its estimation in the calculation of the aforementioned percentile. 

Finally, conditioned on $\epsilon_{0,i}'$, the distribution of the test for large number of samples is given by 
\begin{equation}
    \label{eq:Ddistribution}
    -2\textrm{log}\left.\Lambda\right|_{H_0, \epsilon'_{0}} \overset{p}{\rightarrow} D = \sum_{i=1}^M N_v J_i(g_i)(\epsilon_i - \epsilon_{0,i}')^2,
\end{equation}
which corresponds to a non-central Gaussian quadratic form. Therefore, given a predefined false alarm probability $\alpha$, the test finally reads as
\begin{equation}
    -2\textrm{log}\Lambda\underset{H_0}{\overset{H_1}{\gtrless}} -2\eta = F_D^{-1}(1-\alpha | \widehat{\mathbf{g}}_0), \label{eq:-2eta}
\end{equation}
where $F_D$ is the \ac{CDF} of $D$, which can be obtained by Monte-Carlo simulations or by using some of the approximations given in the literature for Gaussian quadratic forms (see, e.g., \cite{Provost92, Ramirez2019, Moustakas16}). Note that, in \eqref{eq:-2eta}, we have again replaced the true channel values by their estimations. In our proofs, this seems to have a negligible impact on the threshold distribution unless the number of samples is very low (in which case the asymptotic analysis here presented does no longer hold). A summary of the proposed statistical test is provided in Algorithm \ref{alg:Hypo}, where the here presented pointwise comparison is performed along the whole route $\mathbf{P}_c$.

\RestyleAlgo{ruled}
\begin{algorithm}
  \kwTrain{\\
  \For{each $\mathbf{p}\in\mathbf{P}_c$}{
   I. Estimate $\widehat{\mathbf{g}}_{0}$ using \eqref{eq:gestim}\\
   II. Compute $\epsilon'_{0,i}$ for $i=1,\dots,M$ from \eqref{eq:Epsilon0} for a confidence value $\alpha_0$\\
   III. Compute $J(\widehat{g}_{0,i})$ for $i=1,\dots,M$ from \eqref{eq:Ji}\\
   IV. Compute $-2\eta$ using \eqref{eq:-2eta} for a confidence value $\alpha$
   }}
  \kwEval{Received $\mathbf{w}_j$ for $j=1,\dots,N_v$, do\\
  \For{each $\mathbf{p}\in\mathbf{P}_c$}{
   I. Estimate $\widehat{\mathbf{g}}$ using \eqref{eq:gestim}\\
   II. Compute $-2\textrm{log}\Lambda$ using \eqref{eq:GLRTlog}\\
   III. Reject $H_0$ if $-2\textrm{log}\Lambda > -2\eta$\\
   }}
   The point does not belong to $\mathbf{P}_c$ if $H_0$ is rejected $\forall$ $\mathbf{p}$
  \caption{Statistical test for sensing}
  \label{alg:Hypo}
\end{algorithm}

\begin{figure*}[t]
    \centering
         \includegraphics[width=0.9\linewidth]{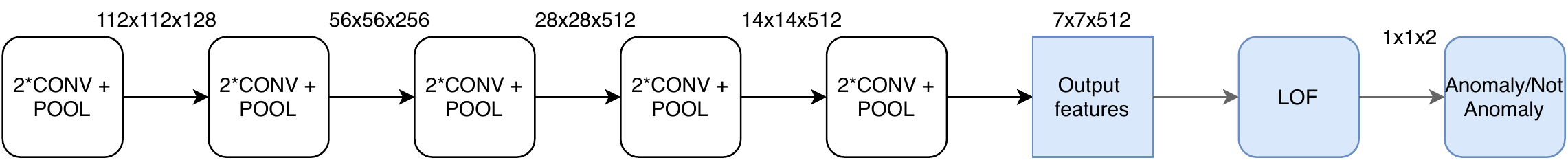}
         \caption{Proposed model. White and blue blocks refer to VGG19 re-used original architecture and to the additional blocks for our task, respectively.}\label{fig:featExt}
\end{figure*}

\section{Machine learning for radio \\ image-based LIS sensing}
\label{sec:MLalgorithm}

In the previous section, we have presented a statistical method to sense the environment based on the received power signal at the different antenna elements of the \ac{LIS}, and hence detecting route deviations from a predefined correct trajectory. This approach exploits the large number of antennas in the \ac{LIS} in the same way as in massive \ac{MIMO} systems. However, the high spatial density of antennas and the large array aperture of \ac{LIS} can be exploited in an alternative way. The basis of this novel technique is using the power of the received signal across the surface as a radio image of the environment, as stated in Sec. \ref{sec:HolographicSensing}. 

\subsection{Model description}
 
  We introduce a \ac{ML} model to perform the anomalous route classification task based on the radio-based images obtained at the \ac{LIS}. The main advantage of this proposal, as we will see, is that it is independent on the data distribution, and no assumptions are needed to its implementation. This is in contrast with section \ref{sec:GLRT} where we considered the noise is Gaussian distributed with noise variance known for the sake of simplicity in the analytical derivations. In reality, these assumptions may not hold. 
 
 In our considered problem, the training data is obtained by sampling the received power at certain temporal instants while the target device is moving along the correct route. In order to reduce both training time and scanning periods, which may be heavy tasks for large trajectories, we resort on transfer learning \cite{pan2009survey}. Because of this matter, the risk of overfitting due to our constraint of short scanning periods is quite significant, being transfer learning also a proper way to tackle it. This allows 
 using a small dataset and therefore improving the flexibility of the system in real world deployments. In our case, we choose the VGG19 architecture \cite{simonyan2014very}.
Due to our specific use case, and the training data constraints, we propose the use of an unsupervised \ac{ML} algorithm named as \ac{LOF} which identifies the outliers presents in a dataset (i.e., the anomalous positions of the target robot) \cite{breunig2000lof}.

The proposed model is detailed in Fig. \ref{fig:featExt}, where, in order to perform the feature extraction, we remove the last \ac{FC} that performs the classification for the purpose of VGG19 and modify it for our specific classification task (anomaly/not anomaly in robot's route).

\subsection{Local Outlier Factor}

\ac{LOF} is an unsupervised \ac{ML} algorithm that relies on the density of data points in the distribution as a key factor to detect outliers (i.e., anomalous events). In the context of anomaly detection, \ac{LOF} was proposed in \cite{breunig2000lof} as a solution to find anomalous data points by measuring the local deviation of a given point with respect to its neighbors.

\ac{LOF} is based on the concept of local density, where the region that compounds a \textit{locality} is determined by its $K$ nearest neighbors. By comparing the local density of a point to the local densities of its neighbors, one can identify regions of similar density, and points that have a substantially lower density than their neighbors (the latter are considered to be outliers). This approach can be naturally applied to the anomalous trajectory deviation detection  as deviated points that are really close to the correct trajectory could be really close in distance, but they would have a lower density compared with the points that actually belong to the correct one, being accurately detected as deviations. Hence, the points belonging to the correct route are used to learn the correct clusters. The strength of the \ac{LOF} algorithm is that it takes both local and global properties of datasets into consideration, i.e., it can perform well even in datasets where anomalous samples have different underlying densities because the comparison is carried out with respect to the surrounding neighborhood. For the reader's convenience, a brief description of the \ac{LOF} theory is provided in the following\footnote{For a more detailed description, the reader is gently referred to \cite{breunig2000lof}.}.

\begin{figure}[t]
    \centering
    \includegraphics[width=0.6\columnwidth]{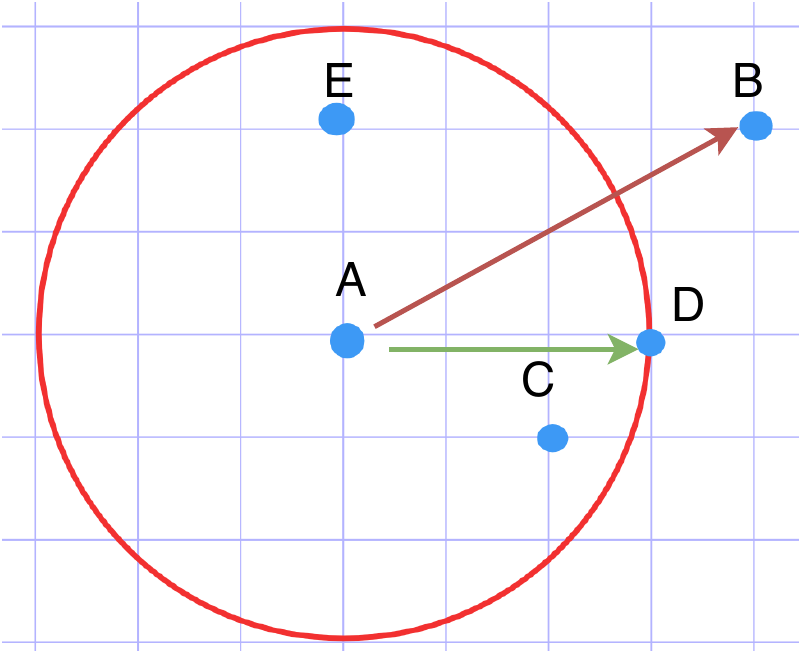}
    \caption{Illustration of Reachability Distance with $K=3$. Manhattan distance used for illustration purposes.}\label{fig:rd}
\end{figure}

The algorithm is based on two metrics, namely the $K$-distance of a point $A$, denoted by $D_K(A)$, and its $K$-neighbors, which is the set $N_K(A)$ composed by those points that are in or on the circle of radius $D_K$ with respect to the point $A$. Note that $K$ is a hyperparameter to be chosen and fixed for computing the clusters. Also note that this implies $|N_K(A)|\geq K$, where $|N_K(A)|$ is the number of points in $N_K(A)$. With these two quantities, the reachability distance between two arbitrary points $A$ and $B$ is defined as
\begin{equation}
    \label{eq:RD}
    RD_K(A, B) = \max\{D_K(A), d(A, B)\},
\end{equation}
where $d(A,B)$ is the distance between points $A$ and $B$. Figure \ref{fig:rd} illustrates the $RD_K$ concept. This means that if a point $B$ lies within the K-neighbors of $A$, the reachability distance will be $D_K(A)=3$ (the radius of the circle containing points $C$, $D$ and $E$), else the reachability distance will be the distance between $A$ and $B$. In the example, $RD_K(A,B)=6$.

Note that the distance measure is problem-specific, being in our case the Euclidean distance between the different features extracted by the VGG19 network. From \eqref{eq:RD}, the \ac{LRD} of a point $A$ is defined as the inverse of the average reachability distance of $A$ from its neighbors, i.e.,
\begin{equation}
    LRD_K(A) = \left(\sum_{B \in N_K(A)}\frac{RD_K(A, B)}{|N_K(A)|}\right)^{-1}.
\end{equation}

According to the \ac{LRD}, if neighbors are far from the point (i.e. more the average reachability distance), less density of points are presented around a particular point. Note this would be the distance at which the point $A$ can be reached from its neighbors, meaning this measures how far a point is from the nearest cluster of points, acquiring low values of \ac{LRD} when the closest cluster is far from the point. This finally give rise to the concept of \ac{LOF}, which is given by
\begin{equation}
    LOF_K(A) = \frac{\sum_{B \in N_K(A)} LRD_K(B)}{LRD_K(A)|N_K(A)|}.
\end{equation}

Observe that, if a point is an inlier, the ratio is approximately equal to one because the density of a point and its neighbors are practically equal. In turn, if the point is an outlier, its \ac{LRD} would be less than the average \ac{LRD} of neighbors, and hence the \ac{LOF} would take large values. In our specific problem, we propose using the \ac{LOF} values to determine whether a point belong to the correct trajectory or from any other point due to a robot deviation.

\begin{figure*}[t]
\centering
\subfloat[Use case scenario. \label{fig:scenario1}]{\includegraphics[width=0.45\linewidth]{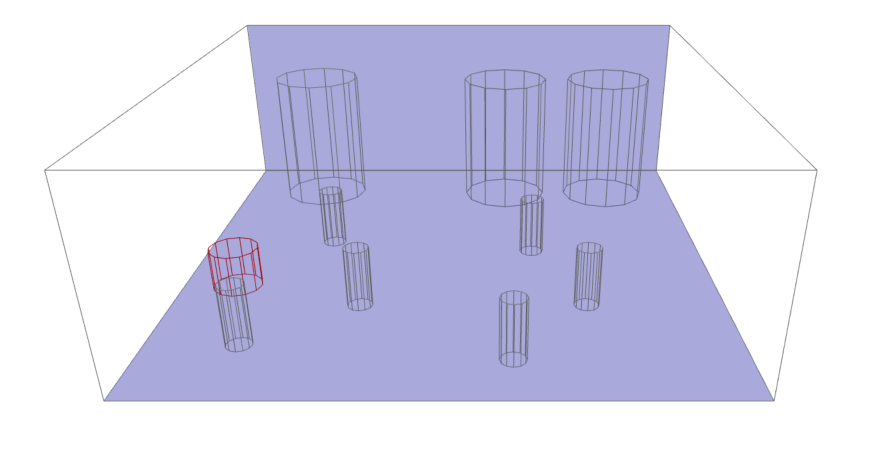}} \hfill
\subfloat[Parallel deviation. \label{fig:scenario2}]{\includegraphics[width=0.25\linewidth]{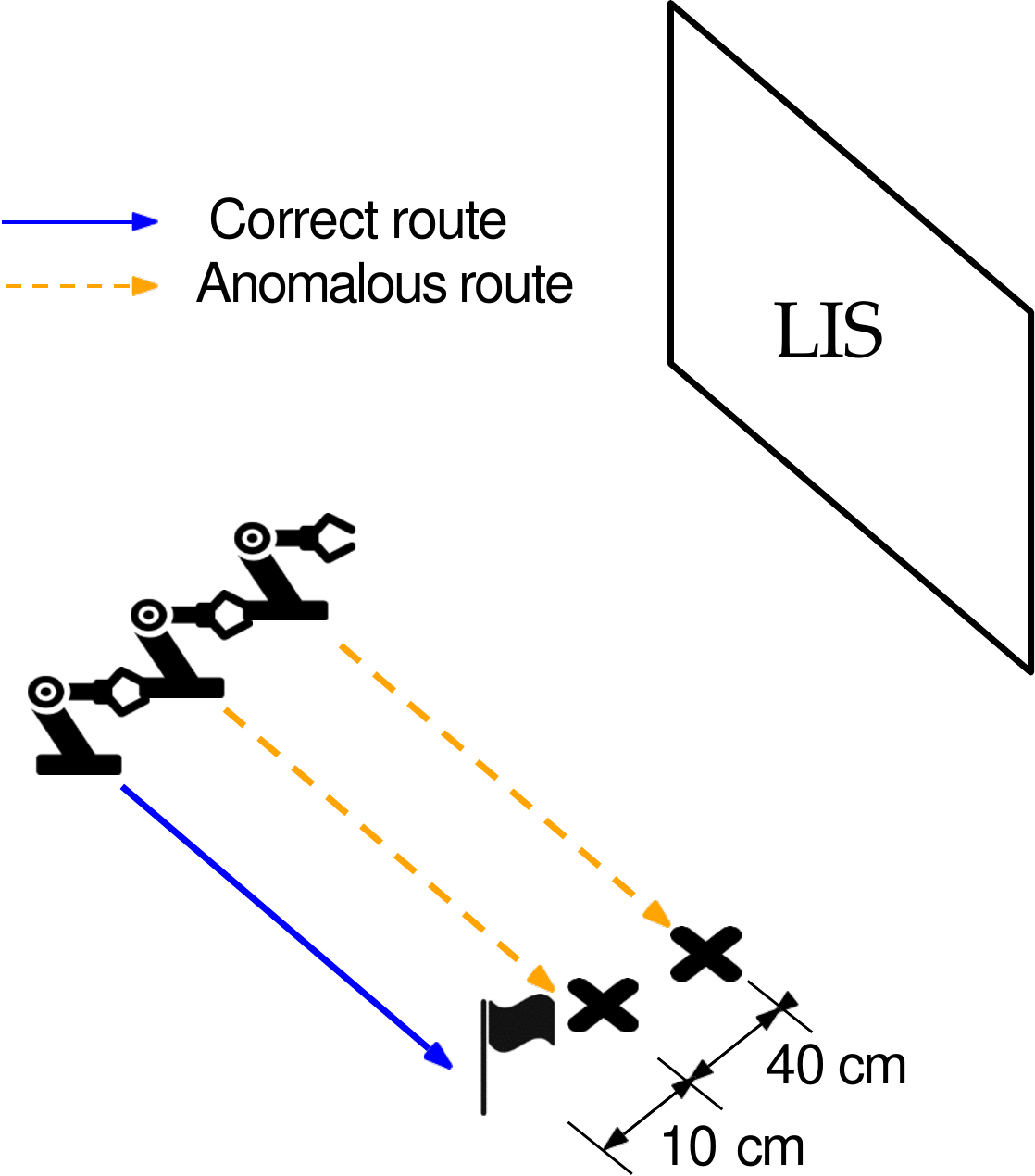}}
\subfloat[Normal deviation. \label{fig:scenario3}]{\includegraphics[width=0.25\linewidth]{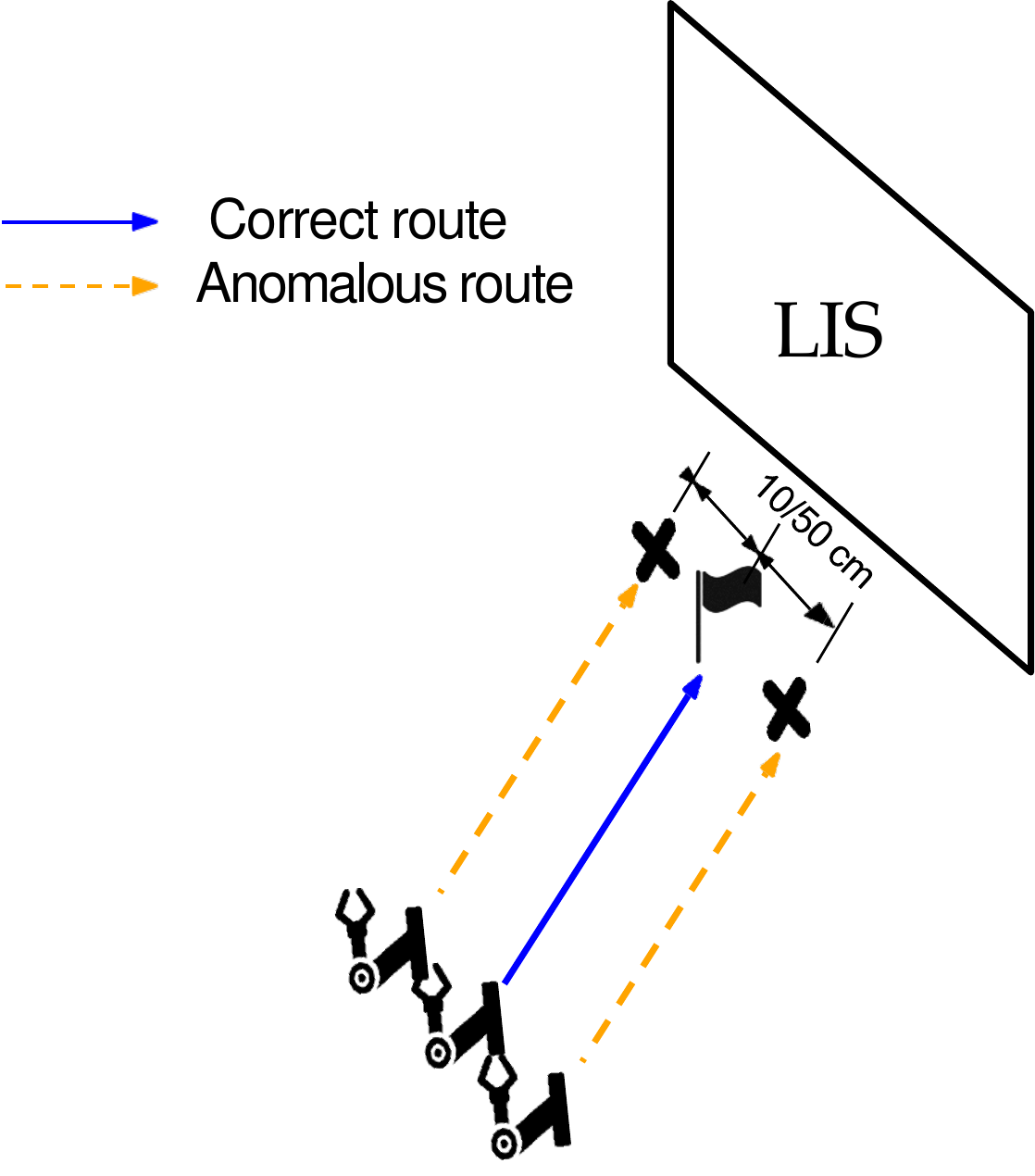}} 
\caption{Simulated scenario.}
\label{fig:scenario}
\end{figure*}

\subsection{Dataset format}
\label{sec:Dataset}

With the algorithm and the model introduced, the remaining component to fully characterize the proposed \ac{ML} solution is the dataset. In our considered problem, the dataset is obtained by sampling the received signal power at each element of the \ac{LIS} while the robot moves along the trajectories. Formally, we can define the possible trajectories (including those composed by both correct and anomalous points) as the set of points in the space $\mathbf{P}_t\in\mathbb{R}^{N_p\times 3}$ being $N_p$ the total number of points in the route. Let us assume the system is able to obtain $N_s$ samples at each channel coherence interval $\forall \;\mathbf{p}_j \in \mathbf{P}_t$, being $\mathbf{p}_j$ for $j=1,\dots,N_p$ an arbitrary point of the route. Hence, the dataset consists of  $T=N_p\times N_s$ samples (monocromatic radio image snapshots of received power). 
Each sample is a gray-scale image which is obtained by mapping the received power into the range of [0, 255]. To that end, we apply min-max feature scaling, in which the value of each pixel $m_{i,j}$ for $i=1,\dots,M$ and $j=1,\dots,N_p$ is obtained as 
\begin{equation}
    \label{eq:pixelmapping}
    m_{i,j} = \ceil* {m_{\textsc{min}} + \frac{(w_{i,j} - w_{\textsc{min},j})(m_{ \textsc{max}}-m_{ \textsc{min}})}{w_{ \textsc{max},j} - w_{ \textsc{min},j}}},
\end{equation}
where $w_{i,j}$ are the elements of $\mathbf{w}_j$, i.e. $w_{i,j}=\|h_{i,j} + n_{i,j}\|^2$, $m_{\textsc{max}} = 255$ and $m_{\textsc{min}} = 0$, and 
\begin{equation}
     w_{ \textsc{max},j}=\max_{\{i=1,...,M\}}\mathbf{w}_{i,j}, \quad  w_{ \textsc{min},j}=\min_{\{i=1,...,M\}}\mathbf{w}_{i,j}
 \end{equation}
  are the maximum and minimum received power value from a point $\mathbf{p}_j$ along the surface.  


The input structure supported by VGG19 is a RGB image of $n_c = 3$ channels. Due to our monocromatic measurements, our original gray-scale input structure is a one-channel image. To solve this problem, we expand the values by copying them into a $n_c = 3$ channels input structure.


Once the feature extraction is performed, the output is $n_c = 512$ channels of size $n_w = 7$ and $n_h = 7$ pixels. Since \ac{LOF} works with vectors, the data is reshaped into an input feature vector formed by $7\times7\times512=25088$ dimensions, meaning our dataset is $\{x^{(i)}\}_{i=1}^T$, where $x^{(i)}$ is the $i$-{th} $n$-dimensional training input features vector (being $n=25088$) and $x^{(i)}_j$ is the value of the $j$-{th} feature.

\section{Model validation}
\label{sec:Validation}

In order to validate both sensing solutions, namely the statistical hypothesis testing and the radio-based image sensing algorithm, we carried out an extensive set of simulations to analyze the performance of the systems in a simple, yet interesting, industrial scenario. 
To properly obtain the received power values, we use a ray tracing software, therefore capturing the effects of the multipath propagation in a reliable way. Specifically, we consider \textsc{Altair Feko Winprop} \cite{winprop}. 

\subsection{Simulated scenario}

The baseline set-up is described in Fig. \ref{fig:scenario1}, a small size industrial scenario of size 484 $m^2$. We address the detection of the deviation of the target robot  (highlighted in red color) in 2 cases: \textit{i)} it follows a fixed route parallel (Fig. \ref{fig:scenario2}), and \textit{ii)} the correct route is normal to the bottom wall, in which the \ac{LIS} is deployed  (Fig. \ref{fig:scenario3}). To evaluate the performance in the detection of anomalies, we consider that the robot may deviate from the correct route at any point, and we test the ability of both systems to detect potential deviations at a distance of, at least, $\Delta d=50$ cm and $\Delta d=10$ cm. These two distances correspond to the cases $ \Delta d \gg \lambda $ and $ \Delta d \approx \lambda $, respectively, denoting $\lambda$ the wavelength.

\begin{table}[h!]
\caption{Parameters}
\label{table_parameters}
\centering
\begin{tabular}{|c|c|c|c|c|c|}
\hline
\begin{tabular}[c]{@{}c@{}}Frequency \\ (GHz)\end{tabular} & \begin{tabular}[c]{@{}c@{}}Tx \\ Power \\ (dBm)\end{tabular} & \begin{tabular}[c]{@{}c@{}}Nray \\ paths\end{tabular} & \begin{tabular}[c]{@{}c@{}}Antenna \\ type\end{tabular} & \begin{tabular}[c]{@{}c@{}}Antenna \\ Spacing (cm)\end{tabular} & \begin{tabular}[c]{@{}c@{}}Propagation \\ model \end{tabular} \\ \hline
3.5                                                        & 20                                                           & 10                                                    & Omni                                                                                                                                                                                            & $\frac{\lambda}{2}$            & Free Space                                                  \\ \hline
\end{tabular}
\end{table}

For the aforementioned cases, we simulate in the ray tracing software $N_p$ points, which correspond to different positions of the robot in both the correct and anomalous routes. Then, $N_s$ radio image snapshots of the measurements are taken at every $\mathbf{p}_j$, $j = 1, \dots, N_p$. The most relevant parameters used for simulation are summarized in Table \ref{table_parameters}.

In our simulations, we set $N_p=367$ and $N_s=10$, thus the dataset is composed of $T=N_p\times N_s=3670$ radio propagation snapshots containing images of both anomalous and non-anomalous situations, as described in Section \ref{sec:Dataset}. Out of $N_p=367$, $N_c=185$ are the snapshots corresponding to the correct route, meaning we have $T_c=N_c \times N_s=1850$ correct data samples, while the remaining are anomalous points. To train the algorithm with the correct points, we split the correct dataset into a 80\% training set 10\% validation set and the remaining 10\% for the test set. During the training phase, the optimum value of $K=3$ (the \ac{LOF} parameter) is obtained by maximizing the accuracy score in the correct validation set. The training procedure was performed in an Intel Xeon machine with 32 CPUs taking around 15 seconds in the offline scanning period.

\subsection{Received power and noise modeling}

The complex electric field arriving at the $i$-th antenna element at sample time $t$, $\widetilde{E}_{i}(t)$, can be regarded as the superposition of each path, i.e.\footnote{Note that the electric field also depends on the point $\mathbf{p}_j$. However, for the sake of clarity, we drop the subindex $j$ throughout the following subsections.}, 
\begin{equation}
    \label{eq:Esum}
  \widetilde{E}_{i}(t) = \sum_{n=1}^{N_r} \widetilde{E}_{i,n}(t) = \sum_{n=1}^{N_r}E_{i,n}(t) e^{j\phi_{i,n}(t)},  
\end{equation}
where $N_r$ is the number of paths and $\widetilde{E}_{i,n}(t)$ is the complex electric field at $i$-th antenna from $n$-th path, with amplitude $E_{i,n}(t)$ and phase $\phi_{i,n}(t)$. From \eqref{eq:Esum}, and assuming isotropic antennas, the complex signal at the output of the $i$-th element is therefore given by
\begin{equation}
    \label{eq:complexSignal}
    y_i(t) = \sqrt{\frac{\lambda^2Z_i}{4\pi Z_0}} \widetilde{E}_{i}(t) + n_i(t),
\end{equation}
with $\lambda$ the wavelength, $Z_0 = 120\pi$ the free space impedance,  $Z_i$ the antenna impedance, and $n_i(t)$ is complex Gaussian noise with zero mean and variance $\sigma^2$. Note that \eqref{eq:complexSignal} is exactly the same model than \eqref{eq:RecSignal}; the only difference is that we are explicitly denoting the dependence on the sampling instant $t$. For simplicity, we consider $Z_i = 1\,\forall\, i$. Thus, the power $w_i(t) = \|y_i(t)\|^2$ is used at each temporal instant $t$ both to perform the hypothesis testing in \eqref{eq:GLRTlog} and to generate the radio image, as pointed out before. Finally, in order to test the system performance under distinct noise conditions, the average \ac{SNR} over the whole route, $\overline{\gamma}$, is defined as\footnote{This is equivalent to averaging over all the points $\mathbf{p}_j$ of the trajectory $\mathbf{P}$.}
\begin{equation}
    \label{eq:snr}
    \overline{\gamma} \triangleq  \frac{\lambda^2}{4\pi Z_0 M T\sigma^2}\displaystyle\sum_{t=1}^{T}\sum_{i=1}^{M} |\widetilde{E}_{i}(t)|^2,
\end{equation}
where $M$ denotes the number of antenna elements in the \ac{LIS}.

\subsection{Noise averaging strategy}

The statistical solution presented in Section \ref{sec:GLRT} has been derived taking into account the presence of noise, and consequently it has implicit mechanisms to reduce its impact in the performance. However, the presence of noise may be more critical in the radio image sensing, since it  impacts considerably the image classification performance \cite{roy2018effects}.

Referring to (\ref{eq:signal}) and (\ref{eq:complexSignal}), since we are considering only received powers, the signal at the output of the $i$-th antenna detector is given by 
\begin{equation}
    w_{i} = \left\|\sqrt{\frac{\lambda^2Z_i}{4\pi Z_0}} \widetilde{E}_{i} + n_i\right\|^2,
\end{equation}
where we have dropped the dependence on $t$.  Also, let us assume the system is able to obtain $S$ extra samples at each channel coherence interval $\forall\; \mathbf{p}_j \in \mathbf{P}_t$. That is, at each point $\mathbf{p}_j$, the system is able to get $N'_{s} = N_s \times S$ samples. Since the algorithm only expects $N_s$ samples from each point, we can use the extra samples to reduce the noise variance at each pixel. To that end, the value of each pixel $m_{i,j}$ is not computed using directly $w_{i,j}$ as in \eqref{eq:pixelmapping} but instead
\begin{equation}
    w'_{i,j} = \frac{1}{S}\sum_{s=1}^{S} w_{i, j, s},
\end{equation}
where $ w_{i, j, s}$ denote the received signal power at each extra sample $s=1,\dots,S$. Note that, if $S\to\infty$, then 
\begin{equation}
    \left. w'_{i,j}\right|_{S\rightarrow\infty} = \mathbb{E}[w_{i,j} | h_{i,j}] = \|h_{i,j}\|^2 + \sigma^2,
\end{equation}


meaning that the noise variance at the resulting image has vanished, i.e., the received power at each antenna (conditioned on the channel) is no longer a random variable. Observe that the image preserves the pattern with the only addition of an additive constant factor $\sigma^2$. This effect is only possible if the system would be able to obtain a very large number $S$ of samples within each channel coherence interval. 

\begin{figure*}[t!]
    \centering
         \includegraphics[width=0.95\linewidth]{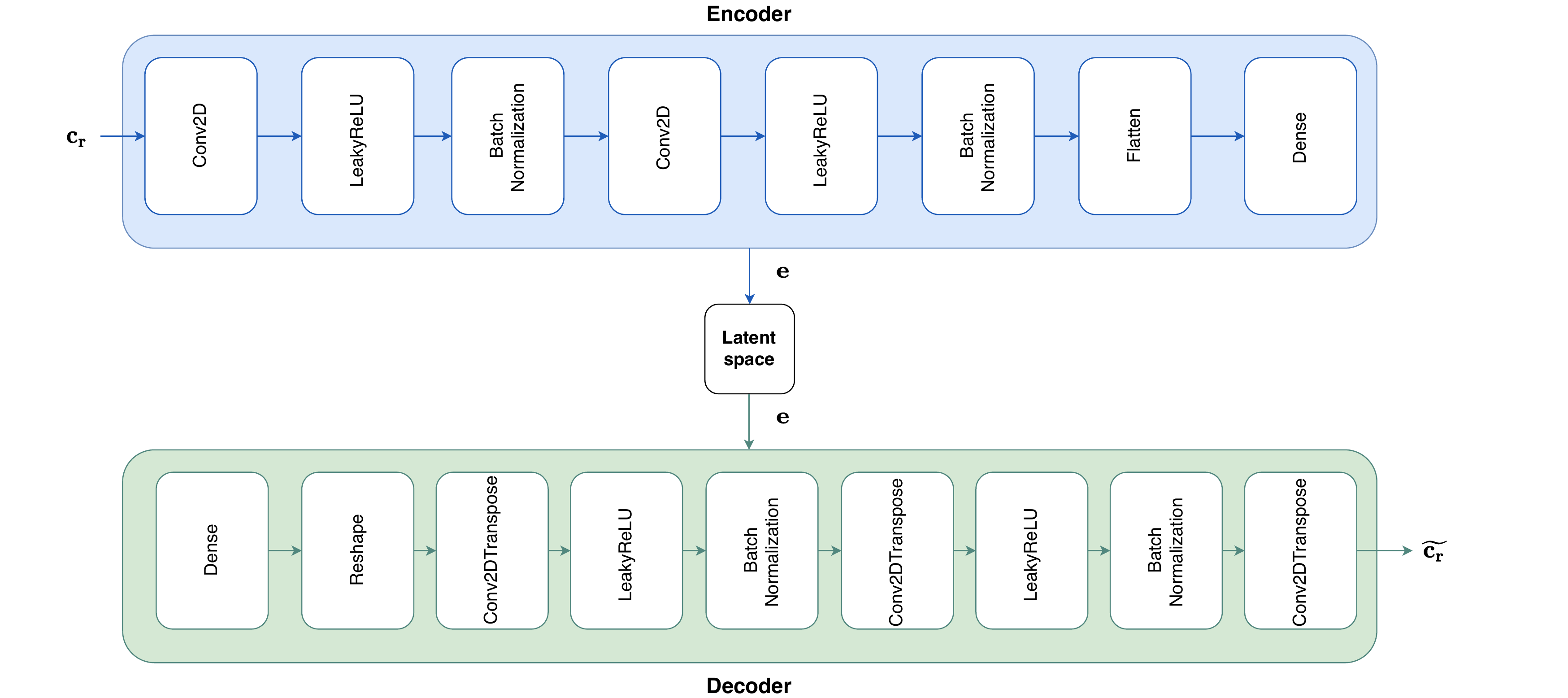}
         \caption{Denoising autoencoder architecture}\label{fig:DAE_architecture}
\end{figure*}

\subsection{Stacked Denoising Autoencoder for image Super-Resolution}
An autoencoder is a type of neural network which tries to learn a representation (denoted as encoding) from the data given as input, usually for dimensionality reduction purposes. Along with the encoding side, a reconstructing side is learnt, where the autoencoder tries to reconstruct from the reduced encoded data a representation as close as possible to its original input. There are several variations of the basic model in order to enforce the learned representation to fulfill some properties \cite{goodfellow2016deep}. Among all of them, we are interested in the \acp{DAE} \cite{vincent2010stacked}. 

The goal of the denoising autoencoder is to reconstruct a "clean" target version  from a corrupted input. In our context, let us assume that we can obtain a target image $\mathbf{t}\in\mathbb{N} \cap [0, 255]^{N}$ and a corrupted input $\mathbf{c}\in\mathbb{N} \cap [0, 255]^{M}$ result of the received power mapping explained in (\ref{eq:pixelmapping}). Also, consider $N\gg M$ and that $\mathbf{t}$ was obtained from a less noisy environment, i.e., the average \ac{SNR} $\overline{\gamma}_t$ is greater than that of $\mathbf{c}$, denoted by $\overline{\gamma}_t \geq \overline{\gamma}_c$. Then, we can perform a resizing of both images such as $R: [0,255]^d \rightarrow \mathbb{R} \cap [0,1]^ R$, meaning we resize both images towards the same dimension $R$ and we normalize the values dividing by $255$, being $\mathbf{t_r} = R(\mathbf{t})$ and $\mathbf{c_r} = R(\mathbf{c})$ the target and corrupted input used to train our \ac{DAE}. Note that, although the two images ($\mathbf{t_r}$ and  $\mathbf{c_r}$) are identical in dimension ($R$ pixels) after the resizing procedure, 
the resolution of the target one is higher because it is obtained in a more favorable scenario (larger \ac{SNR}) and from an initially higher number of pixels $N$. 

To illustrate the approach, a one hidden layer explanation is made for simplicity. The denoising autoencoder can be split into two components: the encoder and the decoder, which can be defined as transformations $\Phi$ and $\Psi$ such that $\Phi: \mathbb{R} \cap [0,1]^R \rightarrow \mathcal{F}$, $\Psi: \mathcal{F} \rightarrow \mathbb{R} \cap [0,1]^R$. Then, the encoder stage of the DAE takes the input $\mathbf{c_r}$ and maps it into $\mathbf{e} \in \mathbb{R}^l = \mathcal{F}$ as
\begin{equation}
    \mathbf{e} = \rho(\mathbf{W}\mathbf{c_r} + \mathbf{b}),
\end{equation}
being $l$ the dimension of the compressed representation of the data, known as the latent space, $\rho$ the element-wise activation function, $\mathbf{W}$ the weight matrix and $\mathbf{b}$ the bias vector. These weights and biases are randomly initialized and updated iteratively through backpropagation. After this process, the decoder stage maps $\mathbf{e}$ to the reconstruction ${\widetilde{\mathbf{c_{r}}}}$ of the same shape as $\mathbf{c_r}$
\begin{equation}
    \widetilde{\mathbf{c_{r}}} = \rho(\mathbf{W'}\mathbf{e} + \mathbf{b'}),
\end{equation}
being $\rho$ the activation function, and  $\mathbf{W'}$ and $\mathbf{b'}$ the parameters used in the decoder part of the network. In our specific case, the reconstruction error, also known as loss, is given by the \ac{MSE} of the pixel values of the target image and the reconstructed image ($R$ pixels), that is\footnote{Note that the summation is made along all the pixel values. However, for the sake of clarity, we drop the subindex in this expression.}
\begin{align}
    \mathcal{L}(\mathbf{t_r}, \widetilde{\mathbf{c_{r}}}) &= \frac{\sum \|\mathbf{t_r} - \widetilde{\mathbf{c_{r}}}\|^2}{R} \notag \\ &= \frac{\sum \|\mathbf{t_r} -   \sigma(\mathbf{W'}\sigma(\mathbf{W}\mathbf{c_r} + \mathbf{b}) + \mathbf{b'})\|^2}{R}.
\end{align}

For the sake of reproducibility, a detailed summary of the proposed architecture is provided in Figure \ref{fig:DAE_architecture}. We have used the Keras \cite{chollet2015keras} library, so the description of the layers corresponds to its notation. The ADAM optimizer with a learning rate $\alpha=0.001$, exponential decay for the 1st and 2nd moment estimates $\beta_1=0.9$ and $\beta_2=0.999$ and $\epsilon=10^{-7}$ have been used for updating the gradient, minimizing the MSE loss function. For the encoder part, Conv2D layers with filter size 64 and 32 have been used, kernel sizes of 3x3 and stride = 2. The activation function LeakyReLU has been used with a slope coefficient $\beta=0.2$. Then, a Batch Normalization layer has been used to maintain the mean output close to 0 and the output standard deviation close to 1. The Flatten layer is used to reshape the output into a vector to feed the Dense layer with a number of neurons $l=16$ which corresponds to the dimension of the latent space. The dimension $l$ was determined by analyzing the learning curves in the training procedure.

In the decoder part, a Dense layer is used again to recover the previous size of the feature vector while the reshaping recovers the initial 2D input structure. Then, Conv2DTranspose layers have been used to perform the reconstruction of the input structure, having an identical configuration than in the encoder side but changing the order of the filters (32 and 64). The LeakyReLU activations and the Batch Normalization are identical. The last layer is composed by 1 filter, kernel size of 3x3 and stride = 1. This last layer is for recovering completely the size as the input structure. Furthermore, the DAE network trains itself to augment the resolution of the input image, because it will remove artifacts resulting from  a lower resolution, by learning from a high resolution target. Then, this reconstructed image will be used for our anomaly detection algorithm. This is advantageous for our problem, leading to a strategy for improving the performance of the system.  

\subsection{Performance metrics}

To evaluate the prediction effectiveness of our proposed method, we resort on common performance metrics that are widely used in the related literature. Concretely, we are focusing on the F1-Score which is a metric based on the Precision and  Recall metrics. First, we need to describe what we consider as a positive or negative event. In our problem, $\text{TP}$ and $\text{FP}$ stand for True and False Positive (anomalous event) while $\text{TN}$ and $\text{FN}$ for True and False Negative (correct event). In this way, the applied metrics are defined as follows:
\begin{itemize}
\item Precision positive (PP) and negative (PN) as the proportion of correct predictions of a given class 
\begin{align}
\text{PP} &=  \frac{\text{TP}}{\text{TP+FP}}, & \text{PN} &= \frac{\text{TN}}{\text{TN+FN}}.
\end{align}
\item Recall positive (RP) and negative (RN) as the proportion of actual occurrences of a given class which has been correctly predicted.
\begin{align}
\text{RP} &=  \frac{\text{TP}}{\text{TP+FN}}, & \text{RN} &=  \frac{\text{TN}}{\text{TN+FP}}.
\end{align}
\item  Positive F1-Score ($PF_{1}$) and Negative F1-Score ($NF_{1}$) as the harmonic mean of precision and recall:
\begin{align}
\text{$PF_{1}$} &=  2\cdot\frac{\text{PP}\cdot \text{RP}}{\text{PP + RP}}, & \text{$NF_{1}$} &= 2\cdot\frac{\text{PN}\cdot \text{RN}}{\text{PN + RN}}. 
\end{align}
\end{itemize}
Note that although the training procedure is fully unsupervised, for our specific evaluation we know the labels of the data samples, meaning we can calculate these metrics, well-known in the supervised learning literature.

\section{Numerical results and Discussion}
\label{sec:Simulations}

We here present some numerical results in order to analyze the performance of both proposals (statistical test and radio image sensing) in our evaluation setup described in Section \ref{sec:Validation}. Generally, in the considered industrial setup, it would be more desirable to avoid undetected anomalies (which may indicate some error in the robot or some external issue in the predefined trajectory) than obtaining a false positive. Hence, all the figures in this section shows the algorithm performance in terms of the $PF_{1}$ metric. 

Also, we mainly focus our results on the radio image sensing algorithm since it is the proposal with a larger number of tunable parameters, whilst the statistical hypothesis testing is used as a benchmark of the \ac{ML} based solution. 

\subsection{Impact of sampling and noise averaging}

First, we evaluate the impact of both available number of samples and the noise averaging technique in the radio image sensing algorithm. To that end,  we consider a \ac{LIS} compounded by $M=32\times32$ antennas and a spacing $\Delta s=\lambda/2$ for a $\Delta d=10$ cm parallel deviation. 

\begin{figure}[t]
    \centering
    \includegraphics[width=\columnwidth]{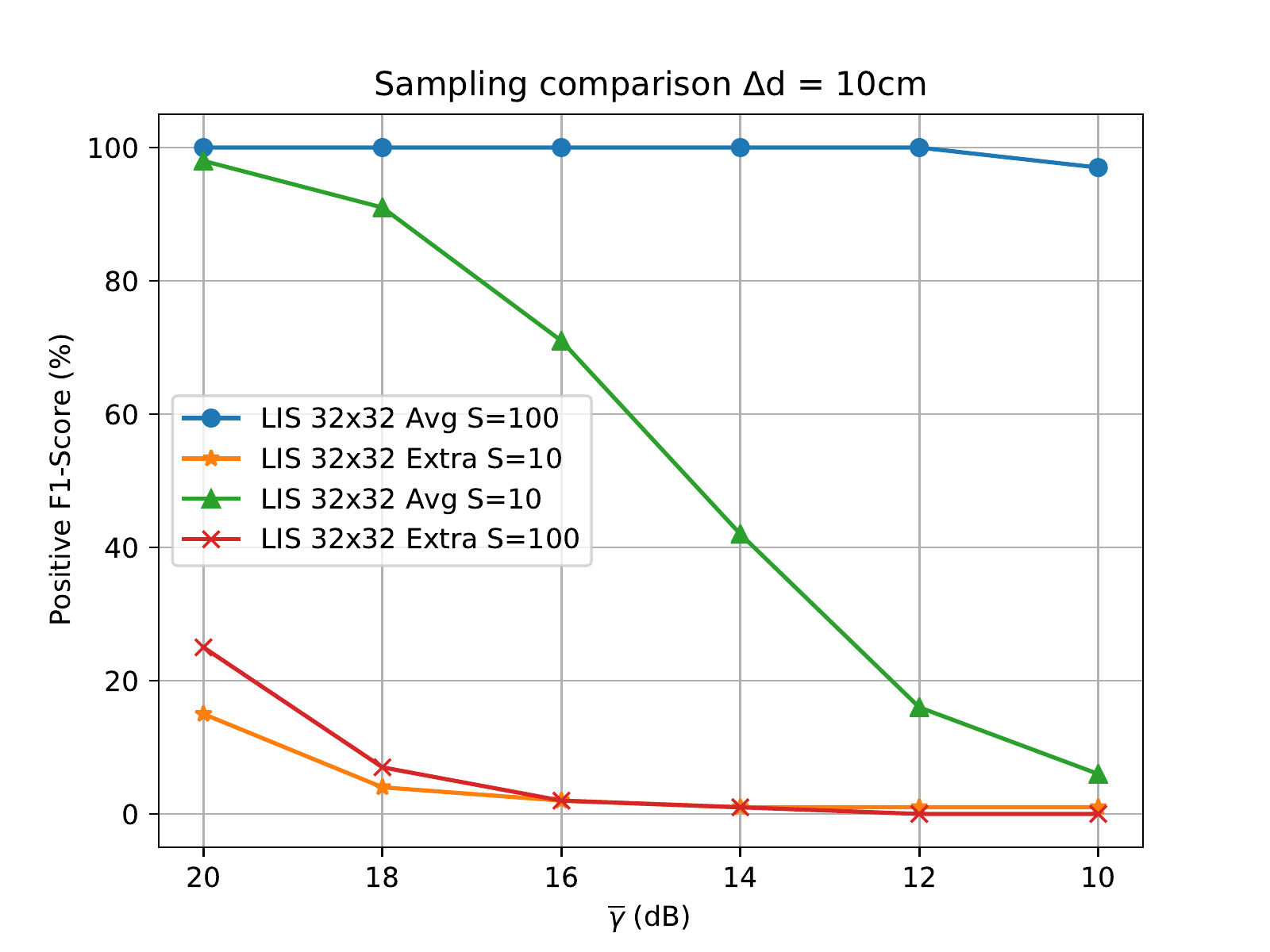}
    \caption{$PF_1$ score for radio image sensing with $M= 32\times 32$ antennas, inter-element distance of $\Delta s = \lambda/2$, correct route parallel to the \ac{LIS}, anomalous points placed at $\Delta d = 10$ cm, and different numbers of samples.}\label{fig:avg_vs_noavg}
\end{figure}

We evaluate two approaches: \textit{i)} using the $S$ extra samples directly as input to the algorithm, being $Tc=N_c\times N_s \times S$, and \textit{ii)} using the $S$ extra samples for averaging. For this particular case, $N_s'\in \{1000, 100\}$. Then $\forall \; \mathbf{p}_j$ we use $S=\frac{N_s'}{N_s}$ samples for obtaining $N_s$ $S$-averaged samples for training the algorithm, being still $Tc=N_c\times N_s=1850$. Note that the number of samples $N_s'$ would depend on the sampling frequency and the second order characterization of the channel, i.e., the channel coherence time and its autocorrelation function. 

Figure \ref{fig:avg_vs_noavg} shows the performance of the system when using $S$ extra samples and $S$ averaged ones respectively. 
As highlighted previously, noise contribution is critical in image classification performance, leading to not achieving a valuable improvement when augmenting the number of samples presented to the algorithm. However, when performing the averaging, results are significantly improved due to the noise variance reduction. As expected, when noise level is higher, more samples are needed to preserve the pattern by averaging, being $N_s'=1000$ the one which yields a better performance. For the following discussions, this sampling strategy will be used, meaning we are using $S=100$ extra samples.

\subsection{Impact of antenna spacing}

The next step is evaluating the impact of inter-antenna distance in the \ac{ML} sensing solution. We fix the aperture to $5.44\times5.44$ m and $S=100$ averaged samples. Then, we assess the performance in both $\Delta d=50/10$ cm for the parallel deviation, and we analyze different spacings with respect to the wavelength ($\lambda/2$, $\lambda$ and $2\lambda$). 

\begin{figure}[t]
    \centering
    \includegraphics[width=\columnwidth]{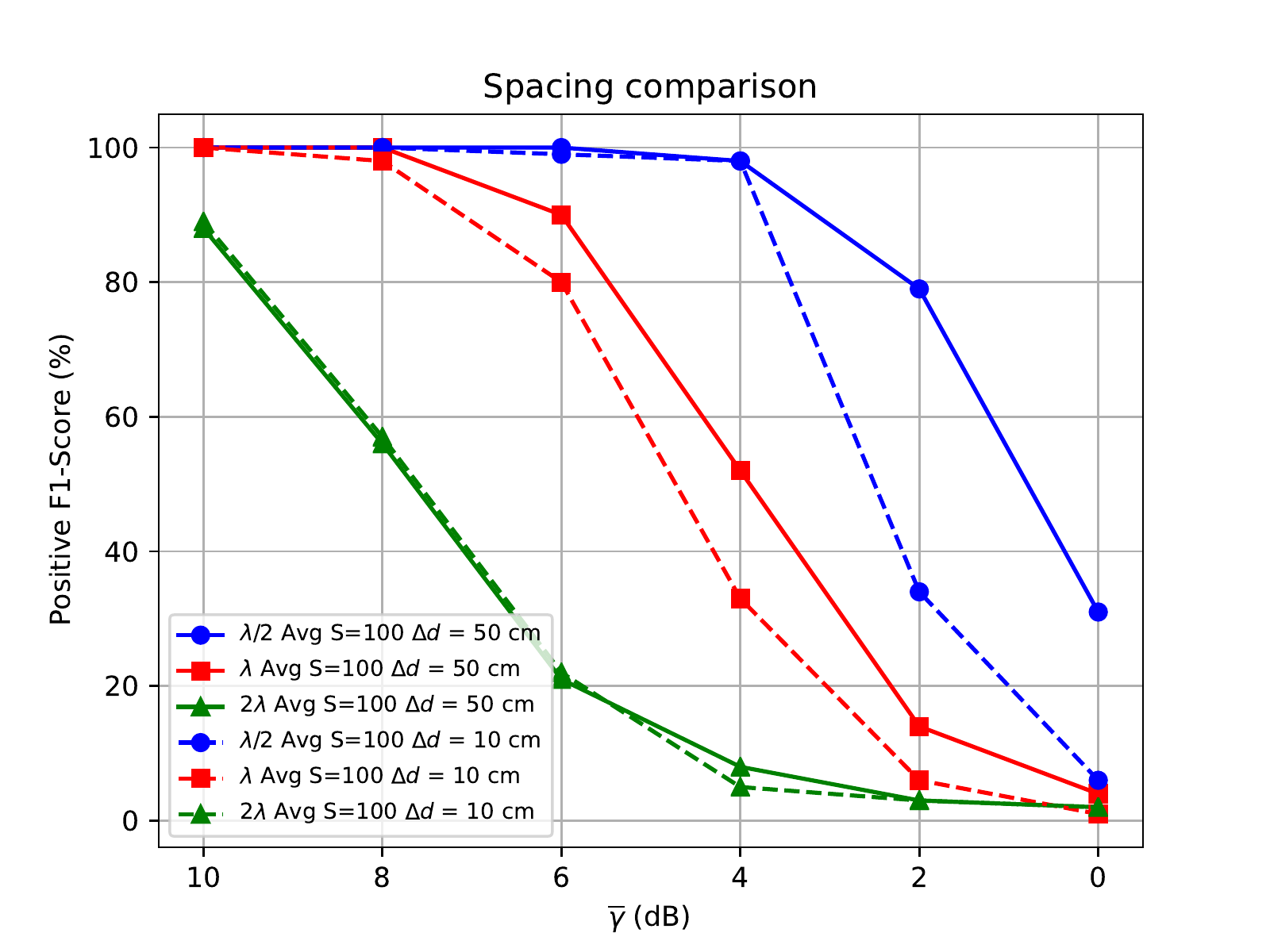}
    \caption{$PF_1$ score for radio image sensing with fixed \ac{LIS} aperture of $5.44\times5.44$ m, correct route parallel to the \ac{LIS}, anomalous points placed at $\Delta d = 10$ and $\Delta d = 50$ cm, $S = 100$ samples, and different number of antennas and spacings.}\label{fig:spacings}
\end{figure}

The performance results for the distinct configurations are depicted in Fig. \ref{fig:spacings}. As observed, the spacing of $2\lambda$ --- which is far from the concept of \ac{LIS} --- is presenting really inaccurate results showing that the spatial resolution is not enough. We can conclude that the quick variations along the surface provide important information to the classifier performance. Besides, this information becomes more important the lower the distance between the routes is. Specially in the range of $\overline{\gamma} \in [10, 4]$ for $\lambda /2$ where the  detection is almost identical regardless of the extra precision needed to detect the deviation when the routes are closer. Furthermore, the effect of antenna densification for a given aperture is highlighted and it can be seen that the lowest spacing leads to the best results.

\subsection{LIS aperture comparisons}

In this case, \ac{LIS} with different apertures have been evaluated. The spacing is fixed to $\lambda/2$, $S = 100$ averaged samples are used while the deviation is $\Delta d = 10$ cm parallel.

 \begin{figure}[t]
    \centering
    \includegraphics[width=\columnwidth]{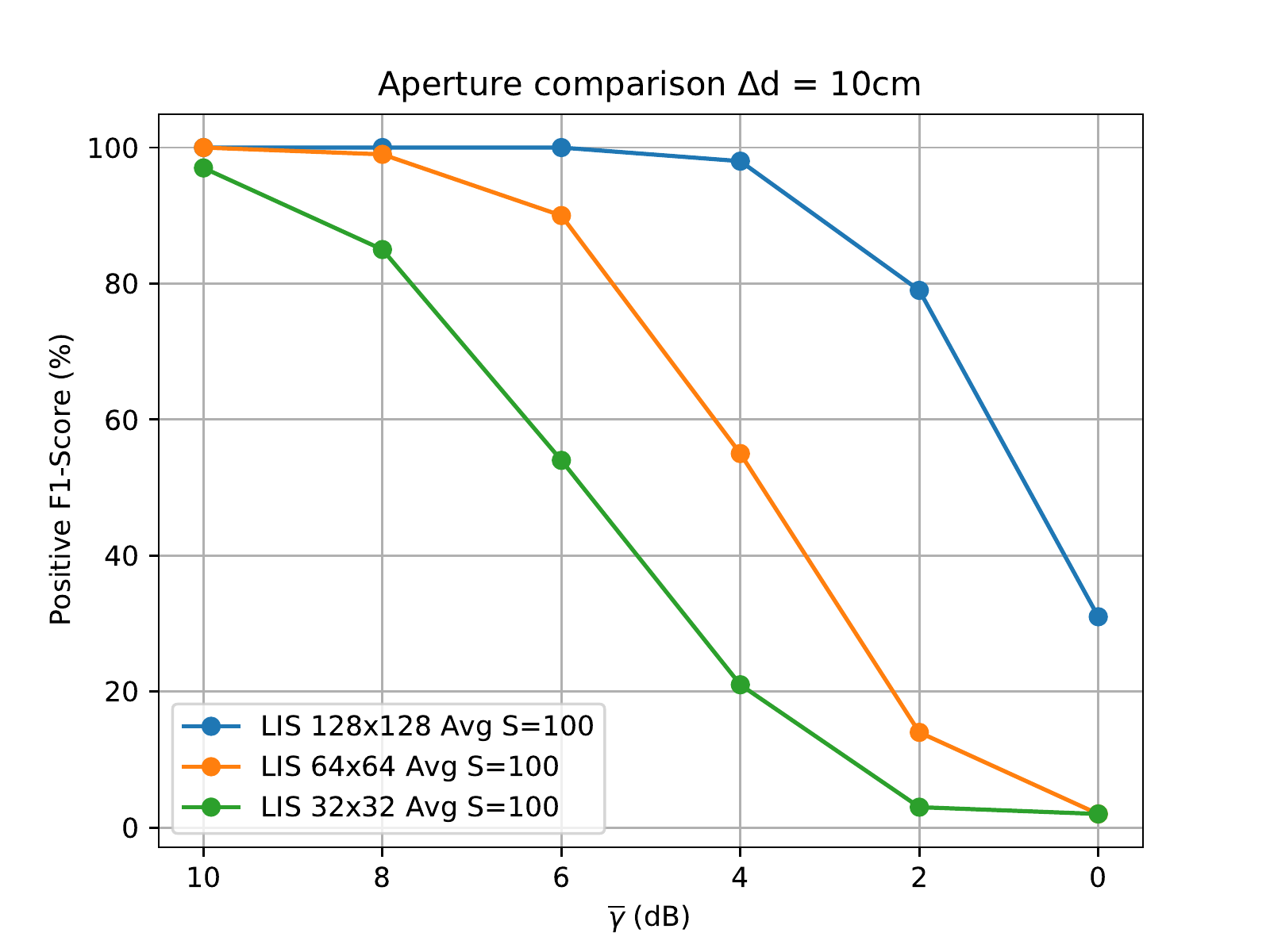}
    \caption{$PF_1$ score for radio image sensing with variable aperture, inter-antenna distance of $\lambda/2$, correct route parallel to the \ac{LIS}, anomalous points placed at $\Delta d = 10$ and $S = 100$ samples.}\label{fig:lambda/2aperturecomparison}
\end{figure}

Looking at Fig. \ref{fig:lambda/2aperturecomparison}, the aperture plays a vital role in the sensing performance. Increasing the number of antennas leads to a higher resolution image, being able to capture the large-scale events occurring in the environment more accurately. Note the usage of incoherent detectors is yielding to a good performance when the aperture is large enough. The key feature for this phenomena is the \ac{LIS} pattern spatial consistency, i.e., the ability of representing the environment as a continuous measurement image.

\subsection{DAE for image Super-resolution evaluation}
In this case, the impact on performance by using DAE is evaluated and compared to the hypothesis test in \eqref{eq:GLRTlog}. We fix the aperture to $M=32 \times 32$, for a parallel deviation of $\Delta d = 10$ cm and an antenna spacing of $\lambda /2$.

 \begin{figure}[t]
    \centering
    \includegraphics[width=\columnwidth]{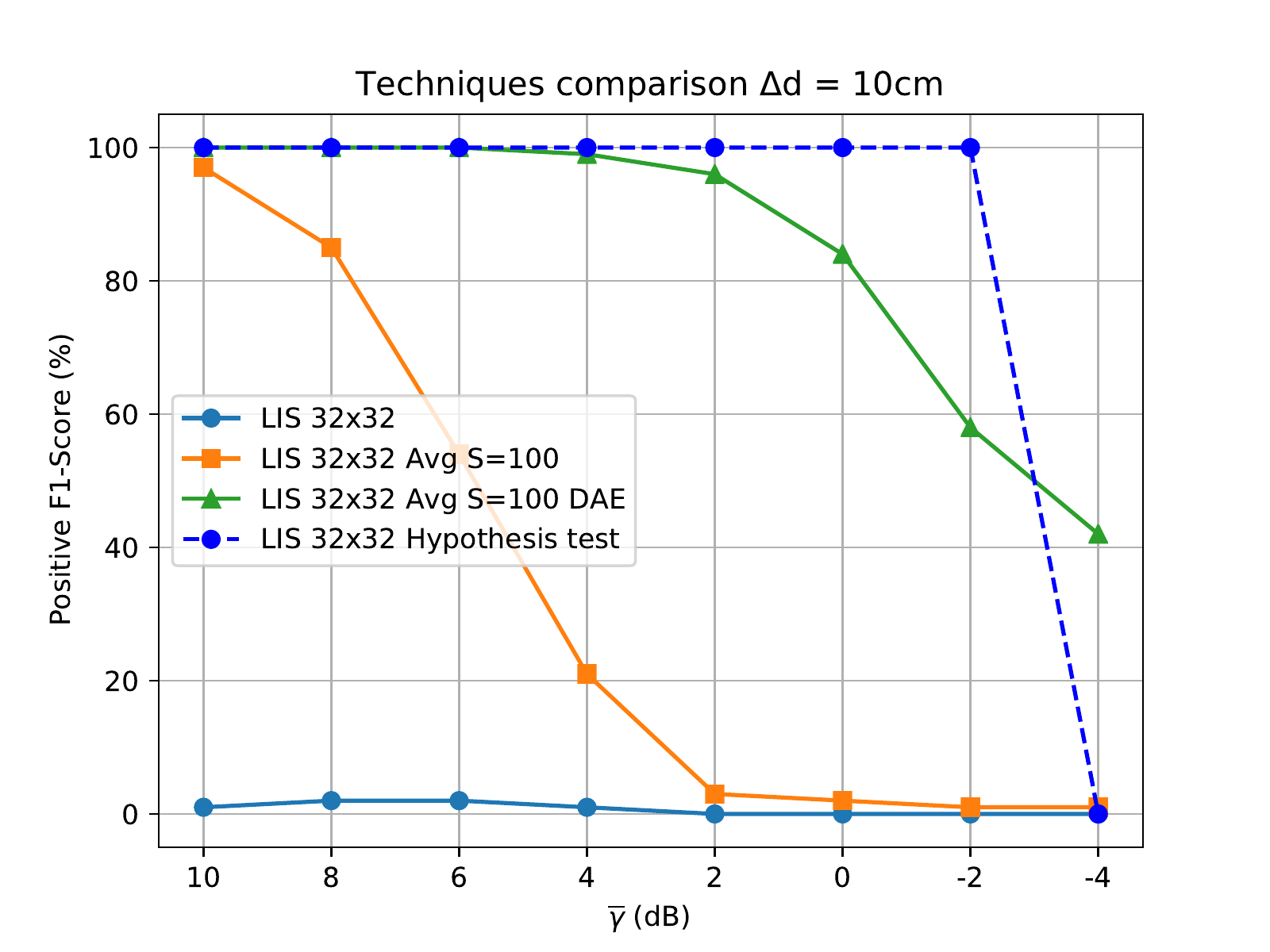}
    \caption{Comparison between radio image sensing and the statistical solution in \eqref{eq:GLRTlog} for $M = 32\times 32$ antennas, correct route parallel to the \ac{LIS}, spacing of $\lambda/2$ and different processing techniques.}\label{fig:techniques}
\end{figure}

For this evaluation, the performance is analyzed in 4 cases: \textit{i)} no pre-processing of images performed, \textit{ii)} $S=100$ averaging strategy, \textit{iii)} image resolution augmentation using DAE, and \textit{iv)} The hypothesis test proposed in Algorithm \ref{alg:Hypo}. For the \ac{DAE}, we assume we have access to a target reference image $\mathbf{t}\in\mathbb{N} \cap [0, 255]^{N}|_{N=128\times 128}$ with $\overline{\gamma} = 10$ dB and our corrupted input is $\mathbf{c}\in\mathbb{N} \cap [0, 255]^{M}|_{M=32\times 32}$ with $\overline{\gamma} \in [10, -10]$ dB. Then, we finally resize both images $R: [0,255]^d \rightarrow \mathbb{R} \cap [0,1]^R|_{R=224\times 224}$, obtaining images of $\mathbf{t_r} \in\mathbb{N} \cap [0,1]^{224\times 224}$ and $\mathbf{c_r} \in\mathbb{N} \cap [0,1]^{224\times 224}$ pixels. 

Regarding Fig. \ref{fig:techniques}, one can see the raw-data (blue line) is yielding to a really poor performance. This is expected taking into account noise can interfere significantly in the local density of the clusters, leading to wrong results. Also, the noise averaging strategy is good enough when noise contribution is negligible, meaning that for improving the results in lower \ac{SNR} scenarios we would need to obtain a higher $S$ which would be unpractical. Finally, the usage of \ac{DAE} for image super-resolution outperforms both methods, allowing to improve the system performance and even work in quite unfavourable \ac{SNR} scenarios. In turn, the hypothesis test derived in Section \ref{sec:GLRT} provides in general the best performance.

However, we must take into account that the statistical test is built based on some key a priori knowledge, namely Gaussian noise with known variance. In the context of estimation, the radio image sensing solution can be seen as a non-parametric approach, which is valid for any baseline distribution and no further assumptions are required. Nevertheless, the performance of the \ac{ML} solution (when \ac{DAE} is employed), presents almost no difference with respect to the ad-hoc test up to $2$ dB of average \ac{SNR}. This is a promising result, since the application of more refined image processing techniques may lead to an increase in performance. Also, note that here we are considering a rather simple scenario, where the scatterers do not move. In a more realistic environment, with the rapid changes in the channel and the temporal dependencies due to the relative positions between users and scatterers in movement, \ac{ML}-based sensing seems a promising solution to learn temporal dependencies in those scenarios where classical solutions become impractical.

\subsection{Route deviations evaluation}
We evaluate now the impact of the separation of deviations and different types of routes in both radio image sensing and the statistical test. To that end, we fix the aperture to $M=32 \times 32$, and a antenna spacing of $\lambda /2$. We will be using all the improvements in the preprocessing of the images to leverage the performance of the \ac{ML} system.

\begin{figure}[t]
    \centering
    \includegraphics[width=\columnwidth]{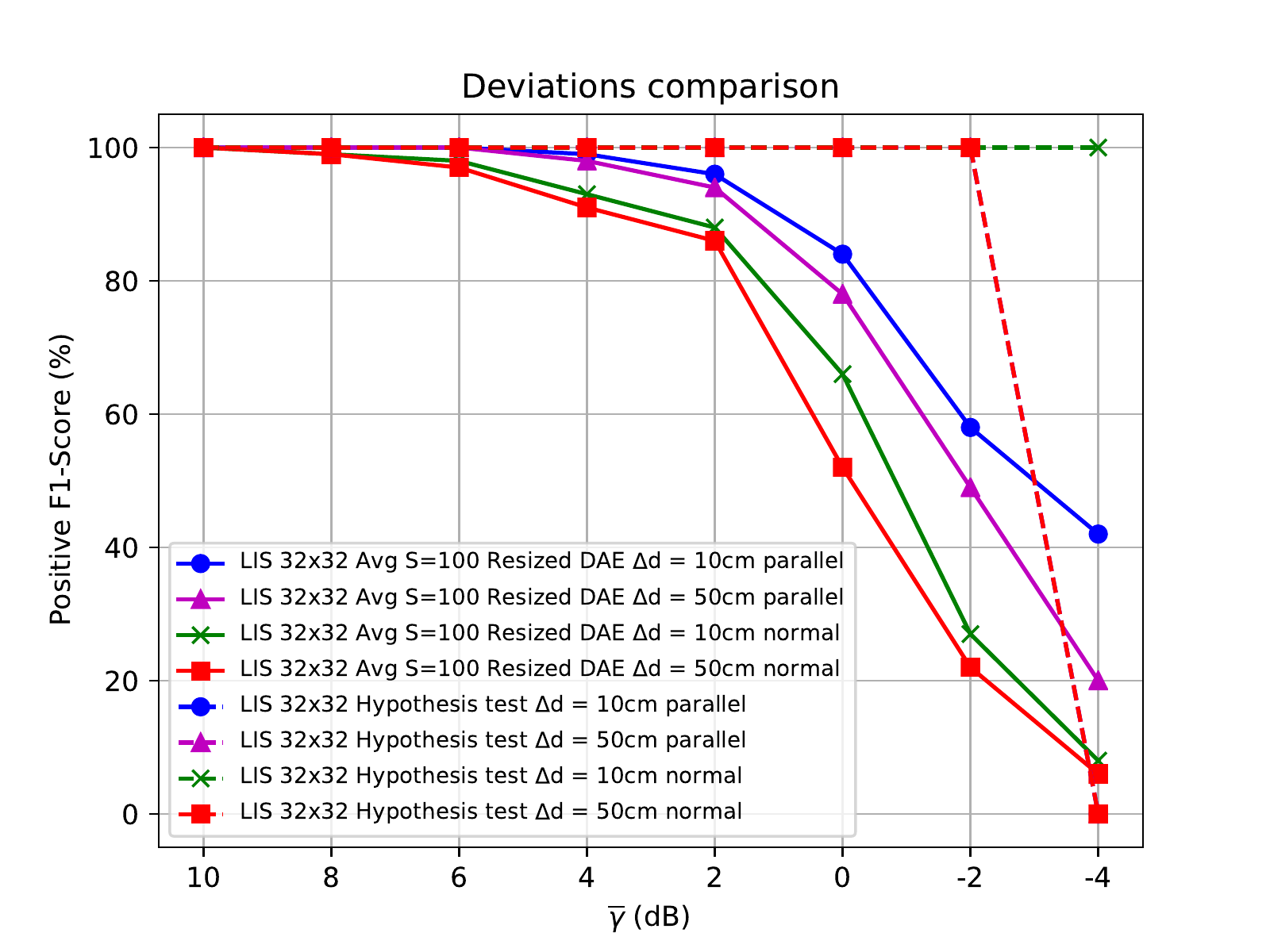}
    \caption{Comparison between radio image sensing and the statistical solution in \eqref{eq:GLRTlog} for $M = 32\times 32$ antennas, different correct routes and spacing of $\lambda/2$.}\label{fig:deviations}
\end{figure}

In Fig. \ref{fig:deviations}, we can see the performance of the system under different deviations and SNR conditions. We can see the system works better the closer the deviation of the routes are. This is an advantage of our proposed approach, the closer the routes are, the more accurate the reconstruction of the DAE is, taking into account the corrupted image $\mathbf{c_r}$ will be more similar to the target image $\mathbf{t_r}$, allowing a better augmentation of the image resolution, so the correct clusters can be learned more accurately. In this way, the \ac{ML} proposed algorithm works better in the cases a standard wireless sensing system would be more prone to failure. Also, the parallel deviations are easier to detect than the normal deviations. The path loss of the points in the parallel routes remains almost identical regardless of the specific point, making it easier to detect. It is important to highlight the SNR definition presented in (\ref{eq:snr}) can influence in the pattern acquisition in the normal deviation cases when points are far from the LIS, which will have a significantly lower instantaneous SNR leading to a more difficult detection.

Note that the abrupt decrease on performance for the hypothesis test is due to the fact that we are using a pointwise test to perform a detection over a whole route (collection of points), as shown in Algorithm \ref{alg:Hypo}. Whilst the \ac{ML} algorithm performs an anomaly detection over the whole correct route, the proposed statistical test is a pointwise comparison, i.e., it checks the validity of the null-hypothesis for each point in the correct route separately. This implies that, in order to detect a point as anomalous, the test has to reject $H_0$ on all the correct points. Consequently, failing in a single point is equivalent to fail in all the points. 

\subsection{Performance evaluation under changing environment}
\begin{figure}[t]
    \centering
    \includegraphics[width=\columnwidth]{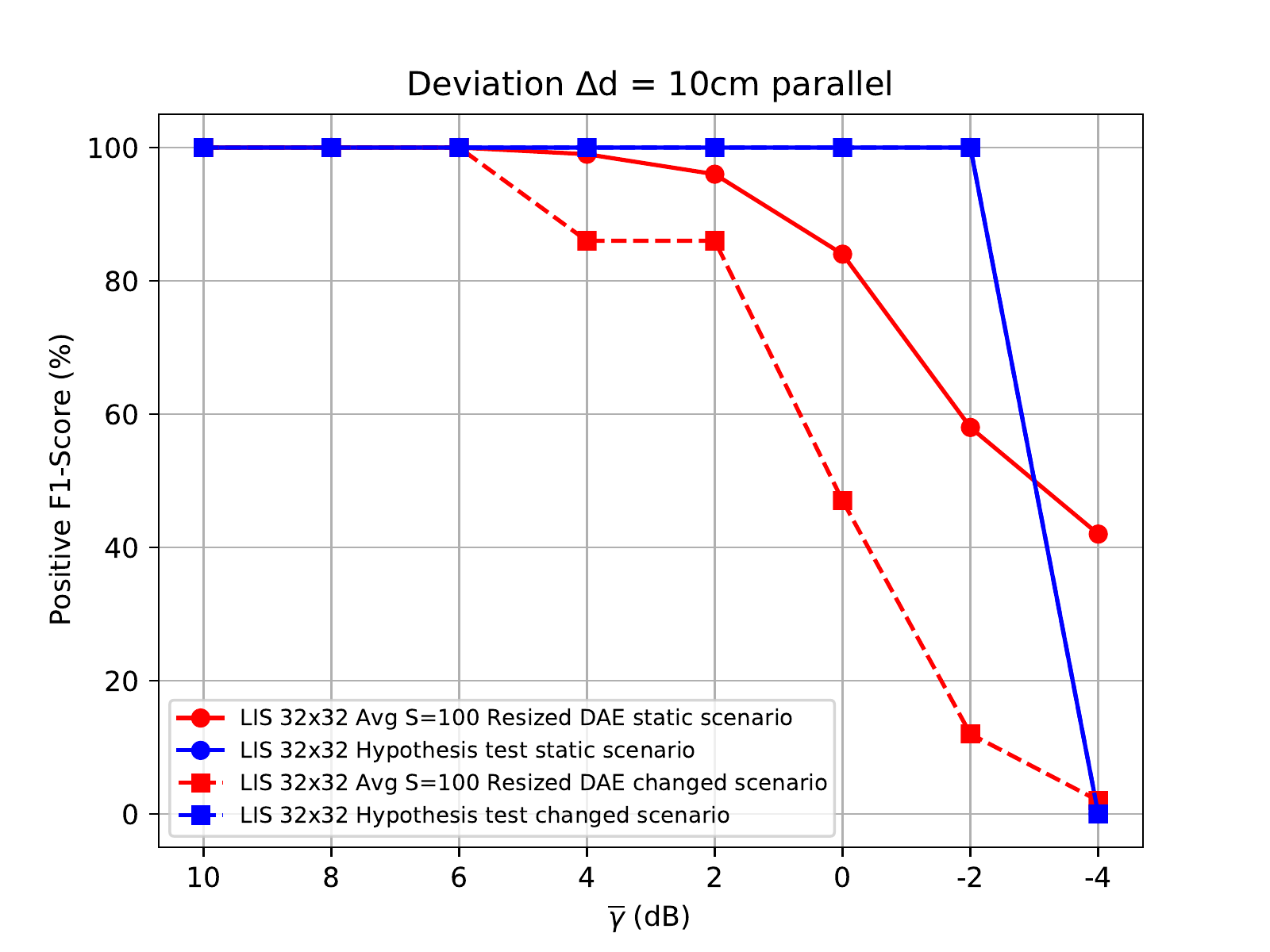}
    \caption{Performance of radio image sensing and the statistical solution in \eqref{eq:GLRTlog} for $M = 32\times 32$ antennas, correct route parallel to the LIS, anomalous points placed at $\Delta d = 10$ and spacing of $\lambda/2$ in a changing environment.}\label{fig:changingenv}
\end{figure}

Finally, we here evaluate the performance of our system when a major change in the scenario occurred, i.e., the relative positions between the scatterers and the transmitter has changed considerably and thus the pattern capture in the radio image no longer matches the original one used in the training phase. Note that, although the considered scenario for testing was assumed to be fixed, we may be interested in extrapolate the performance of the proposed solutions when dealing with environmental changes. To that end, we evaluate the anomaly detection accuracy of both the hypothesis test and the \ac{ML} solution. We fix the aperture to $M=32 \times 32$, for a parallel deviation of $\Delta d = 10$ cm and an antenna spacing of $\lambda /2$. Again, we will use all the improvements in the preprocessing of the images to leverage the performance of the \ac{ML} system. Fig. \ref{fig:changingenv} shows the performance of the system under a changing environment. The hypothesis test is robust to an environmental change, as its performance remains similar as the static case. With respect to the \ac{ML} solution, in the range of $\overline{\gamma} \in [0, -4]$ drops significantly. However, in the proposed scenario, we assume a change in the environment is a really unlikely event, leading to a worsening in the performance in some SNR cases.


\section{Conclusions}
\label{sec:Conclusions}
We have made the first step towards the use of \ac{LIS} for sensing the propagation environment, exploring and proposing two different solutions: \textit{i)} an statistical hypothesis test based on a generalization of the likelihood ratio, and \textit{ii)} a \ac{ML} based algorithm, which exploits the high density of antennas in the \ac{LIS} to obtain radio-images of the scenario. We provide a complete characterization of the statistical solution, and also pave the way to the use of \ac{ML} technique to improve the performance in the second case. As an example, we have shown that the use of denoising autoencoders considerably boosts the performance of the \ac{ML} algorithm. Both proposals are tested in an exemplary industrial scenario, showing that, up to relatively low values of SNR, the performance of the two presented techniques is rather similar. The \ac{ML} solution implies a larger computational effort than the statistical test, but in turn does not require any a priori knowledge, as is the case of the test in which the variance of the noise is assumed in order to reduce analytical complexity. Finally, the results obtained in this system motivate a further study with more complex detectors of I/Q components to  quantify  the  potential  performance  gain  obtained  from  using  I/Q  receivers,  i.e.,  analyzing  the  trade-off between the system complexity and its performance.

%
%
%
%

\bibliographystyle{./IEEEtran}
\bibliography{./biblio}

\end{document}